\pdfoutput=1

\documentclass[
  twocolumn,
  prl,
  showpacs,
  amsmath,
  amssymb,
  superscriptaddress
]{revtex4}

\usepackage{bm}
\usepackage{graphicx}
\usepackage{amssymb}
\usepackage{color}
\usepackage{braket}
\usepackage{empheq}
\newcommand{\tr}{\text{tr}}
\newcommand{\sgn}[1]{\text{sgn}({#1})}
\newcommand*\widefbox[1]{\fbox{\hspace{2em}#1\hspace{2em}}}
\begin{document}

\title{Universal fidelity near quantum and topological phase transitions in finite 1D systems}

\author{E.\ J.\ K\"onig}
\affiliation{Department of Physics, University of Wisconsin-Madison, Madison, Wisconsin 53706, USA}
\author{A.\  Levchenko}
\affiliation{Department of Physics, University of Wisconsin-Madison, Madison, Wisconsin 53706, USA}
\author{N.\ Sedlmayr}
\affiliation{Department of Physics and Astronomy, Michigan State University, East Lansing, Michigan 48824, USA}

\begin{abstract}
We study the quantum fidelity (groundstate overlap) near quantum phase transitions of the Ising universality class in one dimensional (1D) systems of finite size $L$. Prominent examples occur in magnetic systems (e.g. spin-Peierls, the anisotropic XY model), and in 1D topological insulators of any topologically nontrivial Altland-Zirnbauer-Kitaev universality class. The rescaled fidelity susceptibility is a function of the only dimensionless parameter $LM$, where $2M$ is the gap in the fermionic spectrum. We present analytic expressions for the fidelity susceptibility for periodic and open boundaries conditions with zero, one or two edgestates. The latter are shown to have a crucial impact and alter the susceptibility both quantitatively and qualitatively. We support our analytical solutions with numerical data.
\end{abstract}


\maketitle

\textbf{\textit{Introduction.}}
P. W. Anderson's remarkable discovery of the orthogonality catastrophe~\cite{Anderson1967} states that the overlap of two many-body groundstates of two different Hamiltonians, which differ by only a small perturbation, vanishes in the thermodynamic limit; a phenomenon which has recently attracted renewed interest in the expanding research field of quantum information theory.
%P. W. Anderson's remarkable discovery~\cite{Anderson1967}, that the overlap of two many-body groundstates to two different Hamiltonians, vanishes in the thermodynamic limit, recently attracted renewed interest in the expanding research field of quantum information theory.
This branch of quantum physics, which is devoted to the information stored in the wave functions, provides an intriguing arena for both fundamental and applied studies. While one major driving force is the search for a quantum computer, the quantities of interest in quantum information theory, by themselves mathematically fascinating objects, turned out to be useful tools~\cite{ZanardiPaunkovic2006,YangLin2008,AbastoZanardi2008,AmicoVedral2008,Gu2010,DuttaSen2015,ZengWen2015} in the investigation of fundamental phenomena in condensed matter physics, such as quantum phase transitions (QPTs)~\cite{Sachdev2001} and topological phases of matter (TPM)~\cite{Bernevig2013,Shen2013}.

\textbf{\textit{Quantum phase transitions in 1D.}} By definition, a QPT separates two fundamentally different groundstates in the space of externally controllable parameters. Often this `fundamental difference' is the (broken) symmetry of the state. However, the recent advent of TPM lead to the reexamination of this paradigm: here the `fundamental difference' follows from the topological index of the ground state and manifests itself in the appearance of gapless boundary states. In general, a connection between the two concepts of spontaneous symmetry breaking and symmetry protected topological order \cite{ZengWen2015} is not known. However, in one spatial dimension, several archetypical models for QPTs and for TPM are well known to be dual to each other, see Table~\ref{tab:JW}. 

For noninteracting fermions, transitions between distinct TPM are accompanied by a gap closing~\cite{ftn:Nick}. Therefore, the minimal model for such transitions, i.e.~the 1D Dirac Hamiltonian 
%%%%%%%%%%%%%%
\begin{equation}
H = p \tau_x + m \tau_z\,, \label{eq:HDirac}
\end{equation}
%%%%%%%%%%%%%%
turns out to be the universal low-energy theory for topological phase transitions in 1D lattice models of all Altland-Zirnbauer-Kitaev universality classes~\cite{AltlandZirnbauer1997,SchnyderLudwig2008,Kitaev2009}.
In Eq.~\eqref{eq:HDirac} $p$ is the momentum operator and $m$ the mass with $\tau_{x,y,z}$ Pauli matrices.
% \imEJK{Do we need to show this in appendix?}.
Close to criticality the Dirac Hamiltonian is the low energy theory for Ising transitions in two space-time dimensions~\cite{ItzyksonDrouffe1991,GogolinNersesyan2004}.

%%%%%%%%%%%%%%%%%%%%%%%%%%%%
\begin{figure}
\includegraphics[scale=.6]{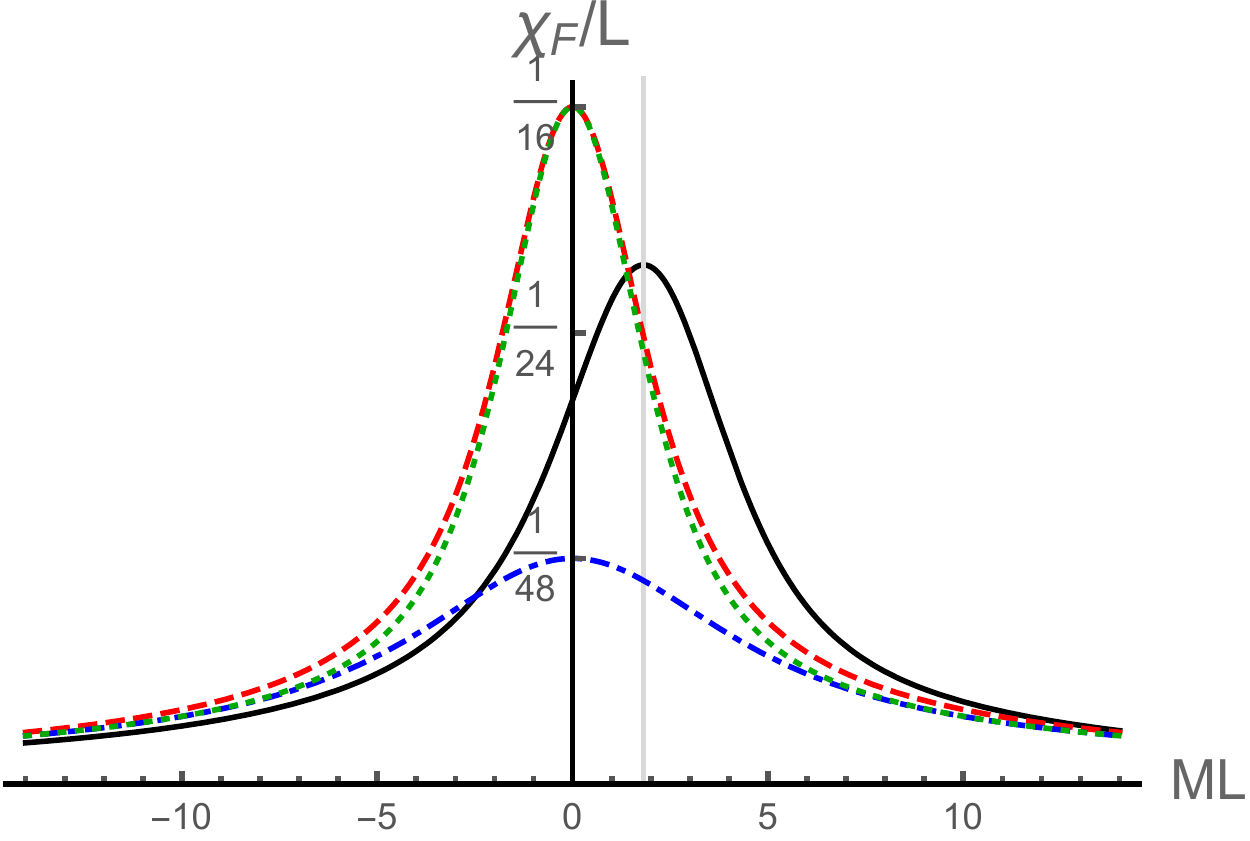} 
\caption{Fidelity susceptibility in finite 1D systems in the case of open boundary conditions [symmetric (asymmetric) mass profile: black, solid (red, dashed)] and closed boundary conditions [periodic (anti-periodic):  blue, dot-dashed; (green, dotted)]. Further explanation is in the main text.}
\label{fig:SuscAllcases}
\end{figure}
%%%%%%%%%%%%%%%%%%%%%%%%%%%%

\textbf{\textit{Fidelity and fidelity susceptibility.}} In this paper we investigate the behavior of the groundstate overlap (or quantum fidelity) near QPTs of the Ising universality class in 1D systems of finite size $L$ and particularly scrutinize the impact of edge states~\cite{SirkerSedlmayr2014}. The fidelity is defined as 
%%%%%%%%%%%%%%%%%%%%	
\begin{equation}
\mathcal F = \vert \braket{\Psi_{0,-}\vert \Psi_{0,+}} \vert ,\label{eq:DefOverlap}
\end{equation}
%%%%%%%%%%%%%%%%%%%%
where $\ket{\Psi_{0,\pm}}$ is the many body groundstate to Eq.~\eqref{eq:HDirac} with mass $m = -M \mp \delta M/2$. In the limit $\vert \delta M L\vert \gg 1$, $\mathcal F$ decays \cite{Anderson1967} at least as a power law, while in the limit $\vert \delta M L \vert \ll 1$ it can be expanded (we will focus on dimension $d = 1$) as
%I(Elio) double checked that it's a power law in Altland Simons book, while in the original article from 1967  Anderson speaks about a power of -1/2.
%%%%%%%%%%%%%%%%
\begin{equation}
\mathcal F \simeq 1 - \frac{(\delta M)^2 L^d}{2} {\chi}_F. \label{eq:DefSusceptibility}
\end{equation}
%%%%%%%%%%%%%%%%

In abstract quantum information theory, the wave function overlap, Eq.~\eqref{eq:DefOverlap}, is related to the Fubini-Study distance $d_{FS}(\ket{\Psi_{0,-}}, \ket{\Psi_{0,+}})$ between elements of the Hilbert space. In contrast, the fidelity susceptibility is related to the Fisher information metric which essentially pulls back the quantum distance $d_{FS}$ to the space of parameters entering the Hamiltonian. It can be related to the real part of the quantum geometric tensor~\cite{ProvostVallee1980,VenutiZanardi2007,KolodrubetzPolkovnikov2013}. 

The experimental relevance of the wave function overlap includes numerous physical systems and phenomena such as the M{\"o}{\ss}bauer effect, X-ray spectroscopy and Kondo physics both in solid state and cold atomic setups. The relationship between the fidelity and the structure factor~\cite{YouLiGu2007}, imaginary time correlation functions~\cite{VenutiZanardi2007}, the probability of excitation after a sudden quench~\cite{DeGrandiPolkovnikov2010}, the scattering matrix~\cite{Ossipov2014} and the spectral function~\cite{GuYu2014}, were uncovered in recent studies. Furthermore, the wave function overlap was shown to enter several observables, such as the average magnetization, for an Ising chain in a quantum field~\cite{RamsDamski2012}.
 All of these relationships will facilitate experimental studies of the fidelity, as they were performed for example in few-body Ising chains~\cite{ZhangSuter2008}.

On the theoretical side, recent years have witnessed outstanding interest in the fidelity close to QPTs, in particular in the context of numerical simulations. One reason is the finite size scaling behavior~\cite{VenutiZanardi2007, SchwandtCapponi2009, Barankov2009, DeGrandiPolkovnikov2010} 
of the fidelity susceptibility, which allows the study of QPTs for which the order parameter is unknown. 
It was proposed in Ref.~\cite{AlbuquerqueCapponi2010}, that
%%%%%%%%%%%%%%%%%%
\begin{subequations}
\begin{equation}
\frac{\chi_F}{L^{2/\nu - d}}  = f_{\chi_F} (L/\xi), \label{eq:scaling}
\end{equation}
%%%%%%%%%%%%%%%%%%
where $\xi = \vert M \vert^{- \nu}$ is the correlation length. For the present case of 1D Dirac fermions ($\nu =1$, $d = 1$), we generalize Eq.~\eqref{eq:scaling} to the case of open boundary conditions for which $\chi_F (M) \neq \chi_F(-M)$
%%%%%%%%%%%%%%%%%%
\begin{equation}
\frac{\chi_F}{L}  = f_{\chi_F} (ML). \label{eq:scaling2}
\end{equation}
\end{subequations}
%%%%%%%%%%%%%%%%%%
This relationship directly follows from the dimensional analysis of Eqs.~\eqref{eq:HDirac}~and~\eqref{eq:DefSusceptibility}. 

Similar but distinct finite size scaling also occurs for various other physical quantities. For example, a different universal function has recently been analyzed in the context of the ground state energy for both closed and open boundary conditions~\cite{GuldenKamenev2016}.

%A universal function in the finite size scaling close to 1D topological QPT also occurs, e.g., in the analysis of the ground state energy, and has been calculated for both closed and open boundary conditions in Ref.~\cite{GuldenKamenev2016}.

%Exact expressions for the fidelity susceptibility of the 1D transverse field Ising model for finite systems with periodic boundary conditions were presented in~\cite{Damski2013,DamskiRams2014}. For Dirac fermions, analytical results for the bulk contribution to the fidelity were also presented in~\cite{MukherjeeSen2012}. In contrast, to the best of our knowledge, only numerical data~\cite{SirkerSedlmayr2014} was reported for systems with open boundary conditions. \imEJK{I had this in the discussions, remove from here.}

\begin{table}
\begin{tabular}{c c c}
1D XY magnet & $\Leftrightarrow$ & 1D fermions \\ 
\hline 
mean coupling $J_x + J_y$ &$\Leftrightarrow$ & hopping $t$ \\ 

anisotropy $J_x - J_y$ &$\Leftrightarrow$ & p-wave pairing $\Delta$ \\ 

transverse magnetic field $h$ &$\Leftrightarrow$ & chemical potential $\mu$ \\ 

staggered coupling $\delta$ &$\Leftrightarrow$ & staggered hopping $\delta$ \\
\hline
$\mathbb Z_2$ symmetry & $\Leftrightarrow $& fermion parity\\
magnetic order &$\Leftrightarrow$& SPT order 
\end{tabular} 
\caption{Comparison of 1D magnetic and fermionic models. In the Kitaev chain (i.e. the anisotropic XY model) $\delta = 0$ while in the SSH and spin Peierls models $J_x - J_y = 0 = h$. For more details, see the main text and Refs.~\cite{Fradkin2013,DuttaSen2015,ZengWen2015,SuppMat}.}
\label{tab:JW}
\end{table} 

\textbf{\textit{Results.}} In this section we present $f_{\chi_F} (ML)$ for four different boundary conditions (see details in Ref.~\cite{SuppMat}). These expressions constitute the major results of this article. 

\textit{Closed boundary conditions.} We first consider the case of periodic boundary conditions (PBC) and antiperiodic boundary conditions (ABC), for which all single particle wave functions obey $\psi(x) = \psi (x + L)$ or $\psi(x) = -\psi (x + L)$, respectively. In these two cases,
%%%%%%%%%%%%%%%%%
\begin{equation}
f_{\chi_F} (ML) = {\frac{\sinh (ML)\mp ML }{16 ML [\cosh (ML)\mp 1] }}, \label{eq:PBCsusc}% \label{eq:ABCsusc}
\end{equation}
%%%%%%%%%%%%%%%%%
where the upper (lower) sign refers to PBC (ABC).
These two results are plotted as a blue dot-dashed and a green dotted curve in Fig.~\ref{fig:SuscAllcases}.

\textit{Open boundary conditions.} We now consider the situation of open boundary conditions, which for Dirac fermions are modelled by means of a ``potential well'' in the mass $m(x)$ entering Eq.~\eqref{eq:HDirac}.
%Recall that, due to Klein tunneling, the scalar potential can not confine Dirac fermions. \textcolor{red}{What is the significance of this here?}
%
We consider two different boundary conditions and introduce them by means of the Su-Schrieffer-Heeger (SSH) lattice model:
%%%%%%%%%%%%%%%%%
\begin{equation}
H_{\rm SSH} =  - t \sum_{j= 1}^{N-1} [ 1+ (-1)^j \delta ] \left (c_{j+1}^\dagger c_j +\textrm{H.c.} \right )\,. \label{eq:SSH}
\end{equation}
%%%%%%%%%%%%%%%%%
Here $t(1\pm\delta)$ is the dimerized hopping between $N$ sites and $c^\dagger_j$ creates a fermion at site $j$.
In the vicinity of criticality ($\delta \rightarrow 0$, $N \rightarrow \infty$), the continuum theory of Eq.~\eqref{eq:SSH} is given by Eq.~\eqref{eq:HDirac} with the identification $-m = M = \delta/a$, where $a$ is the lattice constant and we set the speed of ``light'' $v = 2ta/\hbar \equiv 1$. 
The chain with an even number of sites $N$ is topologically nontrivial (hosts one edge state per boundary) when $\delta >1/(1+N)$ and is topologically trivial when $\delta <1/(1+N)$. In contrast, the chain with an odd number of sites always contains a zero mode which, depending on the sign of $\delta$, is localized on the left or right end of the system.

\begin{figure}
\includegraphics[scale=.3]{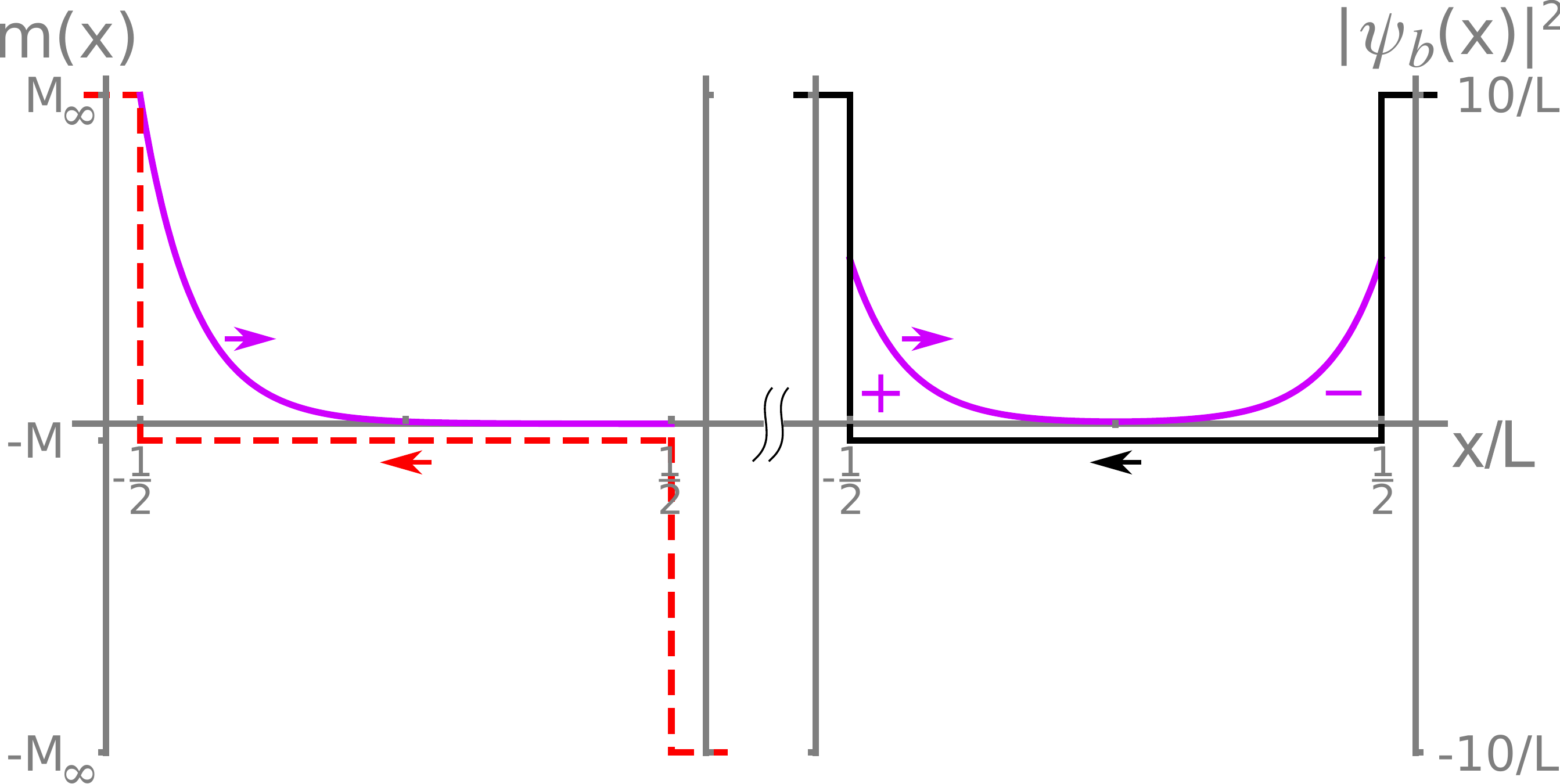} 
\caption{Open boundary conditions for the 1D Dirac model: the asymmetric (left; red, dashed) and symmetric (right; black, solid) mass profile corresponds to, e.g., an SSH chain with odd and even $N$, respectively. The nondispersive wave functions $\psi_b(x)$ (purple) are localized at a single edge for the asymmetric mass profile, while they have equal weight on either boundary in the symmetric case. In the latter case, only the edge state with odd parity contributes to the groundstate fidelity. In this plot $ML= 10$.}
\label{fig:BoundaryCond}
\end{figure}

In the continuum model, these two cases translate to the boundary conditions as follows. Due to the finite system size the wave function has support only in one of the two sublattices. Therefore, one of the two pseudospin projections of the Dirac-spinors vanishes at the system's boundary. We impose this constraint by the following spatial dependence of the mass profile, see Fig.~\ref{fig:BoundaryCond}:
%%%%%%%%%%%%%%%%%
\begin{equation}
m(x) = \begin{cases}M_\infty, & x<-L/2, \\ -M, & -L/2 < x < L/2, \\ \pm M_\infty, & L/2<x.\end{cases} \label{eq:BoundCond}
\end{equation}
%%%%%%%%%%%%%%%%%
The limit $M_\infty \rightarrow \infty$ is to be understood and we refer to the boundary conditions implied by the upper (lower) sign as symmetric (asymmetric). These two general boundary conditions exhaust the possibilities for the open 1D Dirac model. In the symmetric case, which corresponds to the SSH model with an even $N$, edge states appear for $ML >1 $ and are absent otherwise. In contrast, the Callias-Bott-Seeley theorem~\cite{Callias1978,BottSeeley1978} implies the presence of a zero energy state for any value of $ML$ with the asymmetric boundary conditions, as found in the SSH model with an odd $N$.

The fidelity susceptibility for the asymmetric mass profile turns out to be independent on whether the Fermi energy is chosen infinitesimally positive or infinitesimally negative. This is a consequence of the chiral symmetry (or, equivalently, of the particle-hole symmetry)~\cite{SirkerSedlmayr2014,SuppMat}. Therefore, the edgestate formally does not contribute to the fidelity susceptibility, and the result is
%%%%%%%%%%%%%%%%%
\begin{equation}
f_{\chi_F} (ML) =\frac{ML \left[\coth (ML)-2 ML\; \text{csch}^2(ML)\right]+1}{16 (ML)^2}. \label{eq:SSHodd}
\end{equation}
%%%%%%%%%%%%%%%%%
The fidelity susceptibility for the asymmetric boundary conditions is plotted red, dashed in Fig.~\ref{fig:SuscAllcases}.

\begin{figure}
\includegraphics[scale=.6]{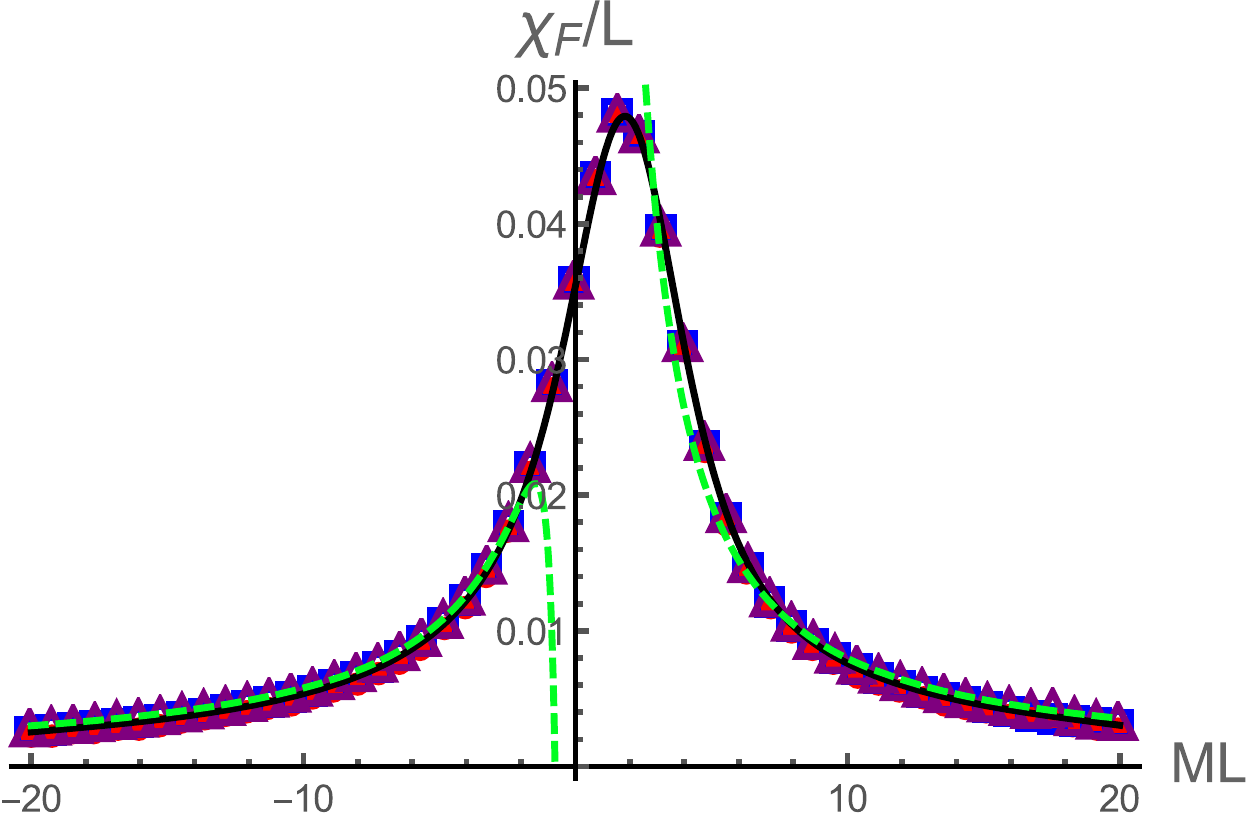} 
\caption{Comparison of analytical and numerical solutions for the fidelity susceptibility. Black, solid: Solution for the Dirac Hamiltonian Eq.~\eqref{eq:HDirac} and symmetric boundary conditions. Green, dashed: Asymptotes, see Eq.~\eqref{eq:SSHevenAsymptotes}. Red dots, blue squares, purple triangles: Numerical solution of the SSH lattice model, Eq.~\eqref{eq:SSH}, for the cases of $N = 100$, $N = 200$ and $N = 400$ respectively.}
\label{fig:ComparisonNumerics}
\end{figure}

Finally, we consider the symmetric mass profile. The fidelity susceptibility in this case is determined by the sum
%%%%%%%%%%%%%%%
\begin{equation}
f_{\chi_F} (ML) = 2 \sum_{\substack{ kL \in \mathcal E^{+}\\ p L \in \mathcal E^{-}}} \frac{k^2 p^2}{\mathcal D_{kL}\mathcal D_{pL} \left (p^2 - k^2\right )^2}. \label{eq:SuscSumMaintext}
\end{equation}
%%%%%%%%%%%%%%%
Here we introduced $\mathcal D_{z} = [z^2 +M^2L^2 - ML]$ and $\mathcal E^{\pm} = \lbrace z \in \mathbb C \vert \tan(z/2) = {z}/[ ML \mp (z^2 +  M^2L^2)^{1/2}]\rbrace$ defines the set of wave numbers associated with even and odd parity states, respectively. The result, Eq.~\eqref{eq:SuscSumMaintext}, can be converted to a closed equation in terms of a two dimensional integral, see Ref.~\cite{SuppMat}. It has asymptotes
%%%%%%%%%%%%%%%%%%%%%
\begin{equation}
f_{\chi_F} (ML) \simeq  \begin{cases} \frac{1}{16 \vert ML \vert}\left ( 1 - \frac{3}{4 \vert ML\vert} \right ), & ML \ll -1, \\ \frac{1}{16 \vert ML \vert}\left ( 1 + \frac{11}{4 \vert ML\vert} \right ), & 1 \ll ML. \end{cases}  \label{eq:SSHevenAsymptotes}
\end{equation}
%%%%%%%%%%%%%%%%%
The fidelity susceptibility is shown as a black curve in Fig.~\ref{fig:SuscAllcases}. We support our analytical results by a numerical calculation for the SSH model, Eq.~\eqref{eq:SSH}, see Fig.~\ref{fig:ComparisonNumerics} and Ref.~\cite{SirkerSedlmayr2014} for more details. It also provides an exemplary proof for the applicability of the critical continuum theory as an approximate description of lattice models. 

%Eventually, we consider the symmetric mass profile. The calculation and result for the susceptibility in this case are much more involved, since the quantization condition for the wave number is determined by a transcendental equation~\cite{GuldenKamenev2016,SuppMat}, which furthermore depends on the gap $M$. In Ref.~\cite{SuppMat} we present this calculation, the analytical solution involves a two dimensional integral. It has asymptotes
%%%%%%%%%%%%%%%%%%%%%%
%\begin{equation}
%\frac{\chi_F}{L} \simeq  \begin{cases} \frac{1}{16 \vert ML \vert}\left ( 1 - \frac{3}{4 \vert ML\vert} \right ), & ML \ll -1, \\ \frac{1}{16 \vert ML \vert}\left ( 1 + \frac{11}{4 \vert ML\vert} \right ), & 1 \ll ML. \end{cases}  \label{eq:SSHevenAsymptotes}
%\end{equation}
%%%%%%%%%%%%%%%%%%
%The fidelity susceptibility is shown as a black curve in Fig.~\ref{fig:SuscAllcases}. We support our analytical results by a numerical calculation for the SSH model, Eq.~\eqref{eq:SSH}, see Fig.~\ref{fig:ComparisonNumerics} and Ref.~\cite{SirkerSedlmayr2014} for more details. It also provides an exemplary proof for the applicability of the critical continuum theory as an approximate description of lattice models. 

%%%%%%%%%%%%%%%%%%%%
%%%%%%%%%%%%%%%%%%%%

\textbf{\textit{Discussion.}} The asymptotic behavior $\chi_F \simeq 1/16 \vert M\vert $ at $\vert ML \vert \gg 1$ and the finite size scaling $\chi_F = L/48 $ at $ ML =0$ for the case of PBC were reported before in different works on the transverse field Ising model~\cite{ZanardiPaunkovic2006,Gu2010} and the SSH model~\cite{SirkerSedlmayr2014}. The functional form of Eq.~\eqref{eq:PBCsusc} is in accordance with Refs.~\cite{Damski2013,DamskiRams2014}. We remind the reader, that periodic and antiperiodic fermionic wave functions occur in the Jordan-Wigner transformed transverse field Ising chains with odd and even number of sites.

For the physical interpretation, one should keep in mind that a large fidelity susceptibility corresponds to a small wave function overlap. In consequence, our calculation shows that, close to the transition, the fidelity is largest in the case of PBC. The technical reason is as follows. For all boundary conditions, the fidelity susceptibility can be represented as a sum over nonzero wavevectors. In the present case ($d \nu < 2$) the sums are ultraviolet convergent and dominated by the infrared. This is because the summand is determined by the overlap of single particle states, which is more susceptible when the ratio between kinetic energy and rest mass is small. The smallest nonzero wavevector for PBC is larger than the smallest wavevector in all other cases, see Ref.~\cite{SuppMat}.

It is noteworthy, how accurately the fidelity for ABC interpolates between the functions for PBC ($\vert ML \vert \gtrsim 5$) and the model with asymmetric mass profile ($\vert ML \vert \lesssim 5$). 
The duality of phases for $M>0$ and $M<0$ implies that in these three cases $f_{\chi_F} (ML)$ is symmetric and peaked at zero, the location of the phase transition. In contrast, for open symmetric boundary conditions, there is no such duality and the reduced fidelity susceptibility is maximal at 
%%%%%%%%%%%%%%%%%%
\begin{equation}
M = M_{c, \chi_F} \equiv b_{\chi_F}/L^\lambda, 
\end{equation}
%%%%%%%%%%%%%%%%%%
with $b_{\chi_F} \approx 1.8$ and $\lambda = 1/\nu =1$ (the shift exponent). For all four cases, Eq.~\eqref{eq:scaling2} implies a bulk-dominated fidelity susceptibility as long as ($ c_{\chi_F}$ of order unity)
%%%%%%%%%%%%%%%%%%
\begin{equation}
\vert M \vert  \gtrsim c_{\chi_F} /L^{\theta}.
\end{equation}
%%%%%%%%%%%%%%%%%%
In the Ising universality class, the rounding exponent is $\theta = 1/\nu = 1$. While our result for the exponents $\lambda$ and $\theta$ conform with the finite size scaling theory of thermodynamic quantities~\cite{FisherFerdinand1967,Henkel2013} the value of $b_{\chi_F}$ is remarkable inasmuch as other observables, such as the groundstate energy~\cite{GuldenKamenev2016} suggest $M_c = 1/L < M_{c, \chi_F}$ for the transition point.
%\imEJK{I couldn't find any work on magnetic susceptibiltiy or specific heat. Need to double check. Otherwise: Future work?}.
As we noted above, $M = M_c$ is the point in parameter space beyond which nondispersive edge states exist. In contrast, at $M \sim M_{c, \chi_F}$ the two edge states decouple, i.e.~their decay length becomes comparable to the system size~\cite{SuppMat}. One should keep in mind that, formally, phase transitions are defined in the thermodynamic limit, in which both $M_{c, \chi_F}$ and $M_c$ approach zero.

A physical intuition for the fidelity susceptibility can be developed on the basis of the groundstate for closed boundary conditions, which is the product state over a collection of two-level systems. At opposite sides of the transitions, the ``pseudo-spins'' tend to be oriented in opposite directions depending on the sign of the mass. Close to the transition the applied ``field'' is weak and thus the ``pseudo-spins'' are more susceptible to changes in the ``field'' and the fidelity smaller ($\chi_F$ larger). Similarly, boundary constraints which are more invasive than PBC generally lead to decoherence of the spin polarization and therefore the fidelity susceptibility is larger. For the symmetric mass profile, however, one should address the two sides of the transition separately. Open boundary conditions imply a node in the dispersing wave functions at the end of, e.g., the SSH chain. Therefore, at the boundary, the form of the wave function is nearly independent of the dimerization and on the topologically trivial side the fidelity in an open system is larger than for PBC. In contrast, on the nontrivial side, the boundary contribution is strongly influenced by edge states. These are susceptible to changes in the dimerization and therefore the boundary contribution to $\chi_F$ is positive. This behavior is reflected in the asymptotes, Eq.~\eqref{eq:SSHevenAsymptotes}.

\textbf{\textit{Conclusion and outlook.}} In this article we derived and discussed the scaling function for the fidelity susceptibility in finite systems close to 1D Ising QPTs. To this end, we employed the critical continuum theory of 1D Dirac fermions subjected to four different types of boundary conditions: open boundary conditions allowing for one or two subgap states as well as periodic and antiperiodic boundary conditions. The fidelity susceptibility close to the transition is smallest for PBC, i.e. in the case of the putatively least invasive constraint. The situation when the boundary conditions allow for two subgap states strongly differs from all others: Only in this case are the two phases not dual  to each other and, as a result, the fidelity susceptibility is not symmetric. The peak value of the reduced fidelity susceptibility defines a critical mass $M_{c, \chi_F} \approx 1.8/L$, which differs from the critical mass obtained by other means~\cite{GuldenKamenev2016}.

Our theory applies to topological phase transitions in which the winding number changes by one. This is because we considered the gap closing of a single Dirac fermion. More generally, a theory of $n$ Dirac fermions applies to certain transitions where the topological winding changes by $n$. In the absence of scattering between different Dirac valleys, the fidelity susceptibility is the sum of the expressions reported in this paper. 
%We emphasize, however, that some transitions with $n >1$ are not described by multiple copies of Eq.~\eqref{eq:HDirac}. For example, the $n = 2$ transition $\Delta \rightarrow - \Delta$ in the Kitaev chain (changing the sign of the anisotropy $J_{x} - J_{y}$ in magnetic language) can involve two distorted Dirac valleys ($\hat P_\pm$ project on different valleys)~\cite{SuppMat}\imEJK{$v(\Delta)$}
%%%%%%%%%%%%%%%%%
%\begin{equation}
%H = \sum_\pm [(\pm \alpha_2 \Delta + p) \tau_x + m(\Delta) \tau_z] \hat P_{\pm}. \label{eq:HDiracDistorted}
%\end{equation}
%%%%%%%%%%%%%%%%%
%In the calculation of many observables, the distortion $\alpha_2 \Delta $ drops out when absorbed into a shift of momenta, but clearly it does not for the overlap of ground states at different values of $\Delta$. 
%Preliminary numerical studies of the Kitaev chain indicate that the fidelity behaves qualitatively different for $\alpha_2 \neq 0$ but recovers the functional form reported in this paper when $\alpha_2 = 0$ (at half-filling). We note that a distortion $\alpha_2 \Delta$ can not occur in an inversion invariant theory for $n = 1$ and leave the study of $n >1$ transitions to future work. 

Since this study was devoted to the universal scaling function of the fidelity susceptibility for 1D Dirac fermions one may wonder about a similar function in higher dimensions. However, such a universal function should not exist, inasmuch as the fidelity susceptibility is expected to depend on the ultraviolet cut-off for $\nu d \geq 2$~\cite{PatelDutta2013,DuttaSen2015}.

Moreover, the generalization of our results to the case of arbitrary $\delta M L$ in Eq.~\eqref{eq:DefOverlap} would reveal further experimentally accessible insights on the orthogonality catastrophe and the role of edge states. Specifically, the overlap function of groundstates on different sides of the transition vanishes for PBC, but not in the case of other boundary conditions~\cite{MukherjeeSen2012}.

Finally, more theoretical work is needed to relate our results for the fidelity susceptibility in finite Ising systems to quantities studied in experiments or by other theoretical methods, such as (boundary) conformal field theory. In particular, the connection between the fidelity and boundary entropy of certain conformally invariant, finite (1+1) dimensional systems was shown in Ref.~\cite{VenutiZanardi2009}. It would be interesting to investigate the Ising critical point in the same spirit. 
%
%Finally, it remains an open question to study the role of edge states for the dynamical counterpart of the fidelity, the Loschmidt echo \cite{SirkerSedlmayr2014,...}.\imEJK{Here is some space if we want to advertise Nick's/our future publication.} \textcolor{red}{...}
%
%
%Also maybe in outlook: Include interaction effects\cite{Sirker2010, others} for the calculation of the fidelity close to topological trnasitions.

%%%%%%%%%%%%%%%%%%%%
%%%%%%%%%%%%%%%%%%%%

%\section{Experimental proposal}
%
%Example SSH: 
%\begin{itemize}
%\item Transition between minima
%\item Susceptibility for dimerization
%\item Measuring process?
%\end{itemize}
%
%SSH in cold atoms \cite{AtalaBloch2013}.
%
%
%\section{Summary from discussion with Nick}
%
%\begin{itemize}
%\item $M_c$ is when $\xi(M) \sim L$.
%\item PBC is lowest, because any boundary conditions tends to decohere polarization
%\item sign of $\chi_B$ in SSH even: start argument from PBC, introduce a cut. Close to boundary w/o boundary states: Wave functions don't change much. w boudnary states: Now there is something to change (which is susceptible).
%\item universtality only for fixed points described by Eq.~\eqref{eq:HDirac}.
%\end{itemize}

\textbf{\textit{Acknowledgements.}} We thank Jia-Hua Gu, Michael Sch\"utt and Kai Sun for useful discussions and acknowledge
hospitality by the Department of Physics and Astronomy
at Michigan State University (A.L. and E.J.K.)
and by the Department of Physics at University of Michigan
(E.J.K.). This work was financially supported in part by NSF Grants No. DMR-1606517 and ECCS-1560732 (E.J.K. and A.L.). Support for this research at the University of Wisconsin-Madison was provided by the Office of the Vice Chancellor for Research and Graduate Education with funding from the Wisconsin Alumni Research Foundation. Support for this research at Michigan State University was provided by the Institute for Mathematical and Theoretical Physics with funding from the office of the Vice President for Research and Graduate Studies.

%%%%%%%%%%%%%%%%%%%%%%%%%%%%%%%%%%%%%%%%%%%%%%

%%%%%%%%%%%%%%%%%%%%%%%%%%%%%%%%%%%%%%%%%%%

\clearpage

%%%%%%%%%% Merge with supplemental materials %%%%%%%%%%
\clearpage

\onecolumngrid

\begin{center}
\textbf{\large Supplemental Material:\\ Universal fidelity near quantum and topological phase transitions in finite 1D systems}
\end{center}
%%%%%%%%%% Merge with supplemental materials %%%%%%%%%%
%%%%%%%%%% Prefix a "S" to all equations, figures, tables and reset the counter %%%%%%%%%%
\setcounter{equation}{0}
\setcounter{figure}{0}
\setcounter{table}{0}
\setcounter{page}{1}
\makeatletter
\renewcommand{\theequation}{S\arabic{equation}}
\renewcommand{\thefigure}{S\arabic{figure}}
\renewcommand{\bibnumfmt}[1]{[S#1]}
\renewcommand{\citenumfont}[1]{S#1}
%%%%%%%%%% Prefix a "S" to all equations, figures, tables and reset the counter %%%%%%%%%%

\section{Jordan-Wigner transformation}
\label{app:JordanWigner}

In this appendix, we review the duality between 1D XY-magnets and the 1D fermionic models for TPM and clarify the notation used in table~\ref{tab:JW} of the main text. 

We consider an XY magnet in a transverse field defined by 
%%%%%%%%%%%%%%%%
\begin{equation}
H_{XY} = - \sum_{j = 1}^{N-1} [J_x^{(j)} \sigma_j^x \sigma_{j+1}^x + J_y^{(j)} \sigma_j^y \sigma_{j+1}^y]  - h \sum_{j = 1}^{N} \frac{\sigma_j^z}{2}.
\end{equation}
%%%%%%%%%%%%%%%%
Here, $\sigma_j^{x,y,z}$ are spin operators (Pauli matrices) at site $j$ and $h $ is the transverse field. For concreteness, we consider the following model of staggered, anisotropic interaction.
%%%%%%%%%%%%%%%%%%%%%
\begin{subequations}
\begin{eqnarray}
J_x^{(j)} &=& J_x [1 + (-1)^j \delta ], \\
J_y^{(j)} &=& J_y [1+ (-1)^j \delta ].
\end{eqnarray}
\end{subequations}
%%%%%%%%%%%%%%%%%%%%%
Using the standard Jordan-Wigner transformation, we rewrite the spin operators in terms of fermionic creation and annihilation operators $c_j^\dagger, c_j$
%%%%%%%%%%%%%%%%%%%%%
\begin{subequations}
\begin{eqnarray}
\frac{\sigma^x_j\pm i \sigma_j^y}{2} &=& \begin{cases} c_j^\dagger e^{-i \pi \sum_{k<j} c_k^\dagger c_k}, \\  c_j e^{i \pi \sum_{k<j} c_k^\dagger c_k} ,\end{cases} \\
\sigma_j^z &=& 2 c_j^\dagger c_j -1.
\end{eqnarray}
\end{subequations}
%%%%%%%%%%%%%%%%%%%%%
Under this transformation, the Hamiltonian becomes
\begin{equation}
H_{XY} = - \sum_{j = 1}^{N-1} [J_x^{(j)} + J_y^{(j)}](c_j^\dagger c_{j+1}+ c_{j+1}^\dagger c_j) +  \sum_{j = 1}^{N-1} [J_x- J_y](c_{j+1}^\dagger c_j^\dagger +  c_j c_{j+1})  - h \sum_{j = 1}^{N} \left (c_j^\dagger c_j - \frac{1}{2}\right ) .
\end{equation}

The identification of the physical meaning of the various terms (hopping $t = J_x + J_y$, p-wave pairing $\Delta = J_x - J_y$, staggered hopping $\delta$ and chemical potential $\mu =  h$) immediately follows.

\section{Fidelity and fidelity susceptibility}

In this appendix, we present details on the derivation of the main result of our paper: The fidelity susceptibility for the Ising transition in 1D described by the Hamiltonian Eq.~\eqref{eq:HDirac}. (Equation numbers without ``S'' refer to the main text.) This appendix contains a section on the symmetry properties of the fidelity susceptibility and the analytical derivation of the results.

%We note that in the context of numerical studies on a lattice with spacing $a$, the following definition for the susceptibility is more common than Eq.~\eqref{eq:DefSusceptibility}
%\begin{equation}
%\mathcal F \simeq 1 - \frac{(\delta M/M)^2 (L/a)^d \tilde\chi_F}{2}.
%\end{equation}
%In the case of the SSH model $\tilde \chi_F/(L/a) = f_{\chi_F} (LM) = \chi_F/L$.

\subsection{Symmetry property of the Fidelity susceptibility}
\label{app:Susc:Symm}

We here show that the fidelity susceptibility evaluated by means of empty states equals the fidelity susceptibility evaluated by all filled states. As a corollary, it follows for the Dirac Hamiltonian, Eq.~\eqref{eq:HDirac}, that the fidelity susceptibility evaluated using all states with energy $E<0$ equals the fidelity susceptibility evaluated using all states with energy $E \leq 0$.

We use the notation 
%%%%%%%%%%%%%%%%
\begin{equation}
\underline A_{p,p'} = \braket{\psi_{(p,M-\delta M/2,-)} \vert \psi_{(p',M+\delta M/2,-)}} \label{eq:overlap}
\end{equation}
%%%%%%%%%%%%%%%%
for the overlap matrix of states with eigenenergy less than or equal to the Fermi energy $E_F$. In our case $\ket{ \psi_{(p,M\pm \delta M/2,-)}}$ are negative-energy eigenstates of Eq.~\eqref{eq:HDirac} with $m = -(M \pm \delta M/2)$.

We rewrite the definition of the fidelity, Eq.~\eqref{eq:DefOverlap}, as follows
%%%%%%%%%%%%%%%%%%%
\begin{equation}
\mathcal F = \sqrt{\det[\underline A \underline A^\dagger]}.
\end{equation}
%%%%%%%%%%%%%%%%%%%
The fidelity susceptibility is thus (the limit $\delta M \rightarrow 0 $ is to be understood in all of the following expressions) 
%%%%%%%%%%%%%%%%%%%%
\begin{subequations}
\begin{eqnarray}
\chi_F &=& - \frac{1}{\delta M^2 L} \tr [\underline A \underline A^\dagger-1] + \frac{1}{2 \delta M^2 L} \tr [(\underline A \underline A^\dagger-1)^2] - \frac{1}{2 \delta M^2 L} \left (\tr [\underline A \underline A^\dagger-1]\right )^2 \label{eq:chiFwithtraces}\\
&=& - \frac{1}{\delta M^2 L} \tr [\underline A^{E> E_F} \underline A^{E> E_F,\dagger}-1] + \frac{1}{2 \delta M^2 L} \tr [(\underline A^{E> E_F } \underline A^{E > E_F,\dagger}-1)^2] - \frac{1}{2\delta M^2 L} \left (\tr [\underline A^{E> E_F} \underline A^{E> E_F,\dagger}-1]\right )^2. \notag \\
\end{eqnarray}
\end{subequations}
%%%%%%%%%%%%%%%%%%%%

The second line follows from the orthonormality and completeness of $\lbrace \ket{\psi_{(p,M\pm \delta M/2,\sgn{E-E_F})}}\rbrace$. Here, we have introduced the overlap matrix $\underline A_{p,p'}^{E>E_F}$ of empty states, defined analogously to Eq.~\eqref{eq:overlap}.

Turning back to our problem of 1D Dirac fermions, we use the chiral symmetry of the Hamiltonian $\lbrace H, \tau_y \rbrace = 0$, to relate all nonnegative energy states to all nonpositive energy states. It follows that the fidelity susceptibility for $E_F = 0^-$ equals the fidelity susceptibility for $E_F = 0^+$. This proves the following assertion from the main text: The single edge state, which is present for asymmetric, open boundary contributions, does not contribute to the fidelity susceptibility.

\subsection{Notation}
Using the Taylor expansion
\begin{equation}
\underline A \simeq \mathbf 1 + \underline A^{(1)} \delta M +\underline A^{(2)} \delta M^2/2,
\end{equation}
Eq.~\eqref{eq:chiFwithtraces} and the fact $\tr[\underline A^{(1)} + \underline A^{(1),\dagger}] = 0$, we rewrite the fidelity susceptibility as follows
%%%%%%%%%%%%%%%%%%%
\begin{subequations}
\begin{equation}
\chi_{\mathcal F} = \chi_{\mathcal F,1} + \chi_{\mathcal F,2},
\end{equation}
where 
\begin{equation}
 \chi_{\mathcal F,1} = \frac{-1}{2 L} \tr[\underline A^{(2)} +\underline A^{(2),\dagger }], \quad
  \chi_{\mathcal F,2} = \frac{1}{2 L} \tr[(\underline A^{(1)})^2 +(\underline A^{(1),\dagger })^2] .
\end{equation}
\label{eq:DefChiF1ChiF2}
\end{subequations}
%%%%%%%%%%%%%%%%%%%%%

\subsection{Closed boundary conditions}
\label{app:sucs:closed}

We first calculate the fidelity susceptibility in the case of periodic and antiperiodic boundary conditions. 

\textbf{Eigenstates.}
We assume $m(x) = -M$ in Eq.~\eqref{eq:HDirac} and periodic boundary conditions (PBC) $\psi(x) = \psi (x + L)$ or antiperiodic boundary conditions (ABC) $\psi(x) = - \psi (x + L)$. After Fourier transform (PBC: $p = 2\pi n/L,\; n \in \mathbb Z$, ABC: $p = \pi (2 n-1)/L,\; n \in \mathbb Z$) we find the following eigenvectors associated to energy $E =  \pm\sqrt{p^2 + M^2}$:
\begin{equation}
\psi_{(M,p,\sgn{E})} (x)= \frac{1}{\sqrt{2}} \left (\begin{array}{c}
- \sgn{E} \sqrt{1 - M/E} \\ 
-\sqrt{1 + M/E}
\end{array} \right ).
\end{equation}

\textbf{Fidelity and fidelity susceptibility.}
Since the discrete wavevectors $k$ are good quantum numbers independently of the mass $M$, the fidelity is the product of the overlap of all negative energy states.% Also note that the overlap of $k = 0$ modes implies that $\mathcal F \propto \theta(M^2 - (\delta M/2)^2)$. Thus, the overlap of groundstates on different sides of the transition is exactly zero~\cite{MukherjeeSen2012} \imEJK{General result by Jia-Hua!}

\begin{eqnarray}
\mathcal F &=& {\prod_k \braket{\psi_{(M- \delta M/2,k,-)} \vert \psi_{(M+ \delta M/2,k,-)}} } \notag \\
&=& \exp\left \lbrace \sum_k \ln [\braket{\psi_{(M- \delta M/2,k,-)} \vert \psi_{(M+ \delta M/2,k,-)}}]\right \rbrace \notag \\
&\simeq & e^{-\sum_k \frac{\delta M ^2 k^2}{8 \left(M^2+k^2\right)^2}} \notag \\
&\simeq & 1 - \frac{\delta M^2 L}{2} \underbrace{\sum_k\frac{k^2}{4 \left(M^2+k^2\right)^2}}_{\equiv \chi_F}.
\end{eqnarray}
We therefore find for PBC
\begin{equation}
\frac{\chi_F}{L} = \sum_{n =-\infty}^\infty \frac{(2\pi n)^2}{4 \left({\bar M}^2+(2\pi n)^2\right)^2} =  {\frac{\sinh (\bar M) - \bar M }{16 \bar M [\cosh (\bar M)-1] }},
\end{equation}
while for ABC we obtain
\begin{equation}
\frac{\chi_F}{L} = \sum_{n =-\infty}^\infty \frac{(\pi(2 n-1))^2}{4 \left({\bar M}^2+(\pi(2 n-1))^2\right)^2} =  {\frac{\sinh (\bar M)+\bar M }{16 \bar M [\cosh (\bar M)+1] }}.
\end{equation}
We here introduced $\bar M = ML$. This concludes the derivation of Eq.~\eqref{eq:PBCsusc} of the main text.

\subsection{Open boundary conditions: asymmetric mass profile}
\label{app:sucs:openodd}

We remark that the Hamiltonian Eq.~\eqref{eq:HDirac} together with the mass profile implied by Eq.~\eqref{eq:BoundCond} has the following behavior under inversion:
\begin{equation}
 \mathcal I \tau_y H \mathcal I \tau_y = H \vert_{M \rightarrow - M}.
\end{equation}
Here $\mathcal I$ inverts $x \rightarrow - x$ and we will classify the states by their parity at $M = 0$.

\textbf{Eigenstates.}
We refer to the following solutions as even states ($p = (2 n-1)\pi/L$)
\begin{equation}
\psi_{(M,p,\sgn{E})}^+(x) =- \frac{(-1)^n}{\sqrt{2 L E^2}} \left[\left(
\begin{array}{c}
  i p \sin (p x) \\
 \left(E+M\right) \cos (p x) \\
\end{array}
\right)+\left(
\begin{array}{c}
- i \left(E-M\right) \cos (p x) \\
 p \sin (p x) \\
\end{array}
\right)\right],
\end{equation}
while odd states ($ p  = 2\pi n/L$) are
\begin{equation}
\psi_{(M,p,\sgn{E})}^- (x) = \frac{(-1)^n}{\sqrt{2 L E^2}} \left[\left(
\begin{array}{c}
 p \cos (p x) \\
 i \left(E+M\right) \sin (p x) \\
\end{array}
\right)+\left(
\begin{array}{c}
 \left(E-M\right) \sin (p x) \\
 -i p \cos (p x) \\
\end{array}
\right)\right].
\end{equation}
Again, $E = \pm \sqrt{p^2 + M^2}$ and $n \in \mathbb N$. The following property can be readily checked:
%%%%%%%%%%%%%%%%%%%%%%%%%%%%
\begin{eqnarray}
\mathcal I \tau_y \psi^{\pm}_{(M,p)}(x) \equiv \tau_y \psi^{\pm}_{(M,p)}(-x)  = \pm\psi^{\pm}_{(-M,p)}(x).
\end{eqnarray}

For the boundary state, it is useful to keep in mind the full space dependence of the mass $m(x)$, even for $\vert x \vert > L/2$, see Fig.~\ref{fig:BoundaryCond}. In this way normalizability imposes the following wave function to be the only zero mode

\begin{equation}
\psi_{(M,0)}(x) = \sqrt{\frac{M}{\sinh(ML)}} \frac{e^{- M x}}{\sqrt{2}} \left (\begin{array}{c}
1 \\ 
-i
\end{array} \right ).
\end{equation}

\textbf{Wave function overlap.}
For the evaluation of the fidelity susceptibtibility we need the wave function overlap, of which we expand diagonal elements up to second order in $\delta M$ and off diagonal elements to first order.

\begin{subequations}
\begin{eqnarray}
\braket{\psi_{(M-\delta M/2,p,-)}^+ \vert \psi_{(M+\delta M/2,p',-)}^+} &\simeq & \left (1-\frac{\delta M^2 p^2}{4 \left(M^2+p^2\right)^2}\right )\delta_{pp'} ,\\
\braket{\psi_{(M-\delta M/2,p,-)}^-  \vert \psi_{(M+\delta M/2,p',-)}^- } &\simeq &\left (1-\frac{\delta M^2 p^2}{4 \left(M^2+p^2\right)^2}\right )\delta_{pp'} ,\\
\braket{\psi_{(M-\delta M/2,0)} \vert \psi_{(M+\delta M/2,0)}} &\simeq & 1- \frac{\delta M^2 L^2}{8} \left(\frac{(ML)^2+1}{(ML)^2}-\frac{1}{\tanh^2(ML)}\right) ,\label{eq:overlapzeros}\\
\braket{\psi_{(M-\delta M/2,p,-)}^+ \vert \psi_{(M+\delta M/2,p',-)}^-} &\simeq & \frac{- 2i \delta M}{\sqrt{p^2+M^2} \sqrt{(p')^2 + M^2} L} \frac{pp'}{(p')^2 - p^2},\\%\braket{\psi_{(M,p)}^- \vert \psi_{(M',p')}^+} &=& \frac{-2i (M' - M) }{EE' L} \frac{pp'}{p^2- (p')^2} \\
\braket{\psi_{(M-\delta M/2,0)} \vert \psi_{(M+\delta M/2,p,-)}^+} &\simeq & (-1)^n\frac{2 i p\delta M \cosh(ML/2)}{\left(M^2+p^2\right)^{3/2}} \sqrt{ \frac{M}{\sinh(ML)}} ,\\
\braket{\psi_{(M-\delta M/2,0)} \vert \psi_{(M+\delta M/2,p,-)}^-} &\simeq & (-1)^n\frac{2 p \delta M\sinh(ML/2)}{\left(M^2+p^2\right)^{3/2}} \sqrt{ \frac{M}{\sinh(ML)}}.
\end{eqnarray}
\end{subequations}

\textbf{Evaluation of fidelity susceptibility.}
Using the above expressions for the wave function overlap we obtain for the case $E_F = 0^-$
%%%%%%%%%%%%%%%%%%%%%%
\begin{subequations}
\begin{align}
\chi_{F,1} &=  \sum_{n \in \mathbb N} \frac{(\pi n)^2}{2((\pi n)^2+{\bar M}^2)^2} = \frac{\coth (\bar M)-\bar M \text{csch}^2(\bar M)}{8 \bar M},  \\
\chi_{F,2} &=  \sum_{{n, m  \in \mathbb N}} \frac{(2n \pi)^2 ((2m-1) \pi)^2}{[(2n \pi)^2 - ((2m-1) \pi)^2]^2} \frac{ (- 8)}{[(2n \pi)^2 + {\bar M}^2][((2m-1) \pi)^2 + {\bar M}^2]}   = \frac{1+{\bar M}[ \text{csch}(\bar M)- \coth \left(\frac{\bar M}{2}\right)]}{16 {\bar M}^2}.\label{eq:chi2SSHodd}
\end{align}
\end{subequations}
The sum of these expressions leads to Eq.~\eqref{eq:SSHodd} of the main text. In the case $E_F  = 0^+$ we include the following additional contribution to $\chi_{F,2}$:
\begin{eqnarray}
\frac{\Delta \chi_{F,2}}{L} &=&- 8 \sum_{n \in \mathbb N} \frac{\bar M}{\sinh(\bar M)} \Big [ \cosh^2\left (\frac{\bar M}{2}\right ) \frac{((2n-1)\pi)^2}{[((2n-1)\pi)^2 + {\bar M}^2]^3} + \sinh^2\left (\frac{\bar M}{2}\right ) \frac{(2n\pi)^2}{[(2n\pi)^2 + {\bar M}^2]^3} \Big] \notag \\
&=& -\frac{\left(-2 {\bar M}^2+\cosh (2 {\bar M})-1\right) \text{csch}^2(\bar M)}{8 {\bar M}^2}. 
\end{eqnarray}
This contribution exactly compensates the effect of the wave function overlap of zeromodes, Eq.~\eqref{eq:overlapzeros}, 
\begin{equation}
\frac{\Delta \chi_{F,1}}{L} = \frac{1}{4} \left (\frac{{\bar M}^2 +1}{{\bar M}^2} - \frac{1}{\tanh^2(\bar M)}\right ) = -\frac{\Delta \chi_{F,2}}{L},
\end{equation}
as required by the general statement, according to which $\chi_F\vert_{E_F =0^-} = \chi_F\vert_{E_F =0^+}$.

\subsection{Open boundary conditions: symmetric mass profile}
\label{app:sucs:openeven}

We first investigate the symmetry properties of the Hamiltonian, Eq.~\eqref{eq:HDirac}, together with the mass profile implied by Eq.~\eqref{eq:BoundCond} under inversion. We find that the Hamiltonian commutes with the following inversion operator:
\begin{equation}
[ H, \mathcal I \tau_z]=0.
\end{equation}
As before, $\mathcal I$ inverts $x \rightarrow - x$. We will classify the states by their parity.

\begin{figure}
\begin{center}
\includegraphics[scale=.5]{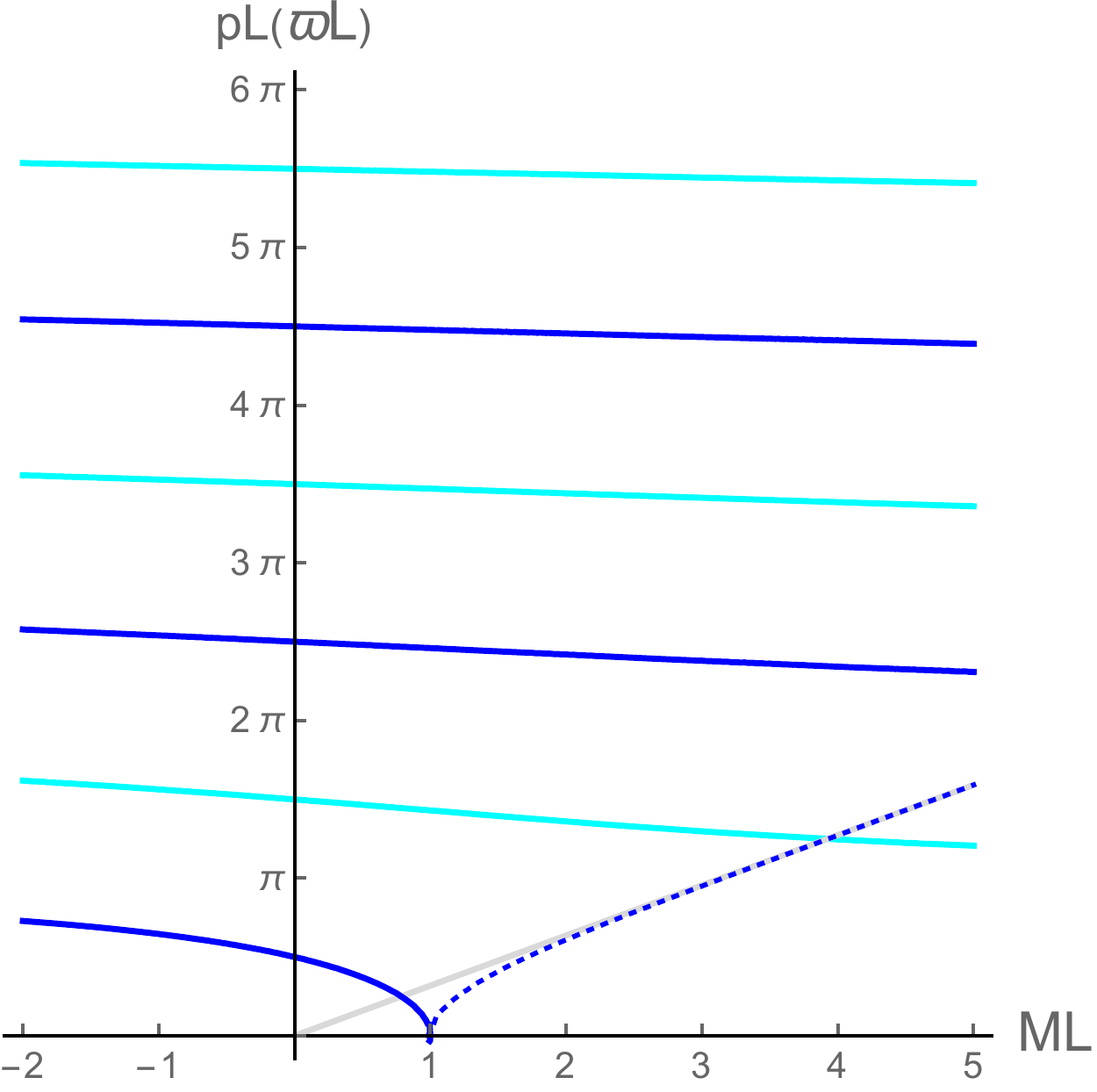} 
\includegraphics[scale=.5]{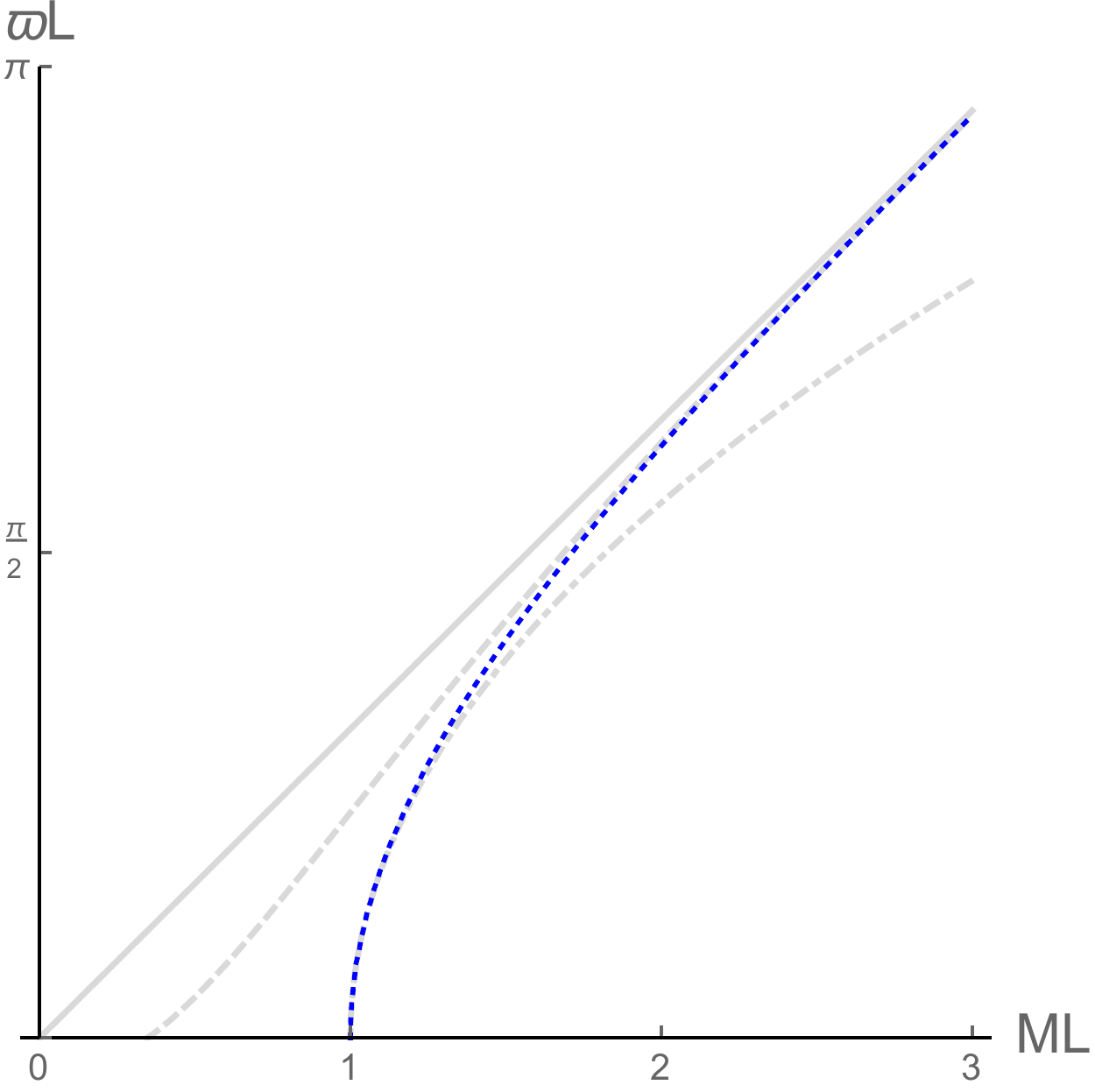} 
\caption{ Left: Spectrum of wave vectors for the even states. Dark blue lines correspond to positive energy, light blue lines to negative energy. For the odd parity states, all states have opposite energy. Note the dotted line, which represents the imaginary wave vector $p = i \varpi$ of the bound state. It approaches the gray asymptote $\varpi = M$ from below. Right: The imaginary solution (blue dotted), together with the asymptotes $\varpi L \simeq \sqrt{3(ML-1)}$ (dotdashed) valid for $ML\rightarrow 1$, $\varpi  \simeq M(1-2e^{-2ML})$ (dashed) and $\varpi = M$ (solid), both valid for large $ML$. }
\label{fig:wavevectors}
\end{center}
\end{figure}

\textbf{Eigenstates.}
The wave functions of even (+) and odd (-) states are 
\begin{subequations}
\begin{equation}
\psi^+_{(M,p,\sgn{E})}(x) = \mathcal N_{(M,p)} \left (\begin{array}{c}
\frac{\cos(px)}{\cos(pL/2)} \\ 
i\frac{ \sin(px)}{\sin(pL/2) }
\end{array} \right ) \text{ and } \psi^-_{(M,p,\sgn{E})}(x) = \mathcal N_{(M,p)} \left (\begin{array}{c}
\frac{ \sin(px)}{\sin(pL/2) } \\ 
i \frac{\cos(px)}{\cos(pL/2)}
\end{array} \right )
\end{equation}
with normalization coefficient
\begin{equation}
\mathcal N^{-1}_{(M,p)}= \sqrt{\int_{-L/2}^{L/2}\left  [\frac{ \sin^2(px) }{\sin^2(pL/2)}+\frac{ \cos^2(px)}{\cos^2(pL/2)}\right ]} = \sqrt{\frac{2L[  M^2 + p ^2 - M/L]}{p^2}}.
\end{equation}
\label{eq:Eigenstates}
\end{subequations}
Note that $\psi^- $ is the chiral partner of  $ \psi^+$, i.e.  $\psi^- = \tau_y \psi^+$, and thus has opposite energy. Direct application of the Hamiltonian, Eq.~\eqref{eq:HDirac}, enforces the energy eigenvalues together with the quantization condition
\begin{equation}
E = \pm (M + p \tan(pL/2) )\text{ and } \frac{p}{M} = \tan(pL).
\end{equation}
The quantization conditions of the different parity eigenstates separately are for even states
\begin{subequations}
\begin{equation}
\tan (pL/2) = \frac{p}{M + E},
\end{equation}
and for odd states
\begin{equation}
\tan (pL/2) = \frac{p}{M - E}.
\end{equation}
\label{eq:quantCond}
\end{subequations}
Each of the even and odd states quantization conditions imply one imaginary solution $p= i \varpi$, with $\varpi >0$, for $ML >1$, see Fig.~\ref{fig:wavevectors}. Physical solutions for real wave vectors have $p >0$, in this case the dispersion relation is $E = \pm \sqrt{p^2 + M^2}$.

%%%%%%%%%%%%%%%%%%%%%%%%%%%
%%%%%%%%%%%%%%%%%%%%%%%%%%%

\textbf{Wave function overlap.}
In view of the inversion symmetry of the problem, the overlap matrix is block-diagonal
\begin{equation}
\underline A _{k,k'} = \left (\begin{array}{cc}
A^+_{k,k'} & 0 \\ 
0 & A^-_{k,k'}
\end{array} \right )
\end{equation}
with overlaps of negative energy eigenstates
\begin{equation}
A^\pm_{k,k'} = \braket{\psi_{(M- \delta M/2,p,-)}^\pm \vert \psi_{(M+ \delta M/2,p',-)}^\pm }.
\end{equation}
Here, the momenta $p\simeq k + \mathcal O(\delta M) $ ($p' \simeq k' + \mathcal O(\delta M)$) obey the quantization condition for a system characterized by mass $M- \delta M/2$ ($M + \delta M/2$). 

The overlap matrix takes the following value for both even and odd states as well as for the case when $k,k' = i \varpi \in i \mathbb R$. 
\begin{equation}
A^+_{k,k'} = A^-_{k,k'} = 4 \mathcal N_{(M- \delta M/2,p)} \mathcal N_{(M+ \delta M/2,p')}\frac{\overline{\delta M} L}{\bar p^2 - \bar{p'}^2} \label{eq:overlapevenodd}
\end{equation}
We need this expression up to $\mathcal O(\delta M^2)$. Keep in mind, that $\bar p= \bar p(\bar M - \overline{ \delta M}/2)$ and $\bar{p'}= \bar{p'}(\bar M + \overline{ \delta M}/2)$. (Barred quantities are $\overline{ \delta M} = \delta M L$ and so on.)

\textbf{Partial cancellation of diagonal and off-diagonal susceptibilities.}
In the calculation of the susceptibility in the case ``SSH-even'' we encounter a partial cancellation of contributions $\chi_{F,1}$ and $\chi_{F,2}$, as defined in Eqs.~\eqref{eq:DefChiF1ChiF2}. We here prove this partial cancellation on general grounds. In the subsequent section, we repeat the proof by explicit calculation, see Eq.~\eqref{eq:SusceptiblitySum}.

Returning to the notation of Eqs.~\eqref{eq:overlap}, we Taylor expand the states and use their normalization to find that 
%%%%%%%%%%%%%%%%%%
\begin{subequations}
\begin{eqnarray}
\underline  A^{(1)}_{p,p'} &=& \braket{\psi_{(p,M,-)} \vert \frac{d}{dM} \psi_{(p',M,-)}}, \\
\underline  A^{(2)}_{p,p'} &=& - \braket{\frac{d}{dM} \psi_{(p,M,-)} \vert \frac{d}{dM} \psi_{(p',M,-)}} .
\end{eqnarray}
\end{subequations}
%%%%%%%%%%%%%%%%%%
In the following, we will use that the matrix $\underline{A}$ is real in the present case. The proof can be readily extended to a more general situation. We further use that odd states are the chiral partners of even states. %and the notation ``$p\,  \text{even}$'' (``$p\,  \text{odd}$'') to denote parity of the wave function (instead of the superscript ``$\pm$'') implicitely by the wave number quantization, Eq.~\eqref{eq:quantCond}.
%%%%%%%%%%%%%%%%%%%
\begin{eqnarray}
\frac{\chi_{F,2}}{L} &=& \frac{1}{L} \sum_{p,k}  \underline  A^{(1)}_{p,k} \underline A^{(1)}_{k,p}  \notag \\
&=& - \frac{1}{L} \left [\sum_{p,k }\braket{ \frac{d \psi_{(p,M,-)}^+}{dM} \vert\psi_{(k,M,-)}^+}\braket{\psi_{(k,M,-)}^+ \vert \frac{d \psi_{(p,M,-)}^+}{dM} } + \sum_{p,k}\braket{ \frac{d\psi_{(p,M,-)}^-}{dM}  \vert\psi_{(k,M,-)}^-}\braket{\psi_{(k,M,-)}^- \vert \frac{d \psi_{(p,M,-)}^-}{dM}} \right ] \notag \\
&=& - \frac{1}{L} \left [\sum_{p,k }\braket{ \frac{d \psi_{(p,M,-)}^+}{dM} \vert\psi_{(k,M,-)}^+}\braket{\psi_{(k,M,-)}^+ \vert \frac{d \psi_{(p,M,-)}^+}{dM} } + \sum_{p,k}\braket{ \frac{d\psi_{(p,M,+)}^+}{dM}  \vert\psi_{(k,M,+)}^+}\braket{\psi_{(k,M,+)}^+ \vert \frac{d \psi_{(p,M,+)}^+}{dM}} \right ] \notag \\
&=&  \frac{1}{L} \left [2 \sum_{p,k}\braket{ \frac{d \psi_{(p,M,-)}^+}{dM} \vert\psi_{(k,M,+)}^+}\braket{\psi_{(k,M,+)} ^+\vert \frac{d \psi_{(p,M,-)}^+}{dM} }-\sum_{p,k; \xi,\zeta = \pm }\braket{ \frac{d \psi_{(p,M,\xi)}^+}{dM} \vert\psi_{(k,M,\zeta)}^+}\braket{\psi_{(k,M,\zeta)}^+ \vert \frac{d \psi_{(p,M,\xi)}^+}{dM} } \right ] \notag \\
&=& \frac{2}{L}  \sum_{p,k}\braket{ \frac{d \psi_{(p,M,-)}^+}{dM} \vert\psi_{(k,M,+)}^+}\braket{\psi_{(k,M,+)} ^+\vert \frac{d \psi_{(p,M,-)}^+}{dM} } - \frac{\chi_{F,1}}{L}. \label{eq:partialcancellation}
\end{eqnarray}
%%%%%%%%%%%%%%%%%%%

In the very last line, we used the resolution of identity in the space of symmetric wave functions. We thus see, that contributions from diagonal parts of the matrix $\underline A$ partially cancel the contributions from the off-diagonal part. We now proceed with the explicit calculation.

\textbf{Diagonal contribution to fidelity susceptibility.}
We expand the diagonal part of Eq.~\eqref{eq:overlapevenodd} to second order in $\delta M$. It is important to keep in mind that all wave vectors are a function of the mass $\bar p = \bar p (M)$.
%To this end we use the following properties of the pseudo momenta
%\begin{subequations}
%\begin{eqnarray}
%\bar p(\bar M) &=& \bar M \tan (\bar p) \\
%\bar p'(\bar M) &=& - \frac{\bar p}{\mathcal D_{\bar p}} \\
%\bar p''(\bar M) &=& - \frac{2( \bar M -1) \bar p (\bar p^2 + {\bar M}^2)}{\mathcal D_{\bar p}^3} \\
%\bar p'''(\bar M) &=& \frac{2 \bar p (\bar p^2 + {\bar M}^2)(\bar p^4 - 2 \bar p^2 (3 - 4{\bar M} + {\bar M}^2) -3 ({\bar M}^2 - {\bar M})^2)}{\mathcal D_{\bar p}^5}. 
%\end{eqnarray}
%\end{subequations}
We obtain

\begin{subequations}
\begin{equation}
A_{k,k} = 1- \frac{(\delta M L)^2}{24} \left [\sum_{n = 1}^4 \frac{2 A_{n}(\bar M)}{\mathcal D^n_{\bar k}} \right ]
\end{equation}
where $\mathcal D_{\bar k} = \bar k^2 + \bar M^2 - \bar M$ and
\begin{eqnarray}
A_{1} (\bar M)&=& 2,\\
A_{2} (\bar M)&=& -(9-20\bar M+4\bar M^2)/2,\\
A_{3} (\bar M)&=& -4{\bar M}(3-5\bar M+2\bar M^2),\\
A_{4} (\bar M)&=& -6(\bar M^2-\bar M)^2.
\end{eqnarray}
\end{subequations}
%This leads to the following expression for $\chi_{F,1}$
%\begin{equation}
%\chi_{F,1} = \frac{L}{12}   \left [\sum_{n = 1}^4  A_{n}(\bar{M})\left ( s_n(\bar{M}) -\frac{1}{(\bar{M}^2-\bar{M})^n}\right )\right ]. \label{eq:chiF1Sol}
%\end{equation}
%We here introduced the sums

For the analytic solution we solve the following sums
\begin{equation}
s_n(\bar M) = \sum_{\bar k \in \mathcal E \backslash \lbrace 0 \rbrace} \frac{1}{(\bar k^2 + \bar M^2-\bar M)^n}, \quad \mathcal E = \lbrace \bar k \in \mathbb C \vert f(\bar k) \equiv \tan(\bar k) - \bar k/\bar M = 0 \rbrace.
\end{equation}
by contour integration leading to the following final expression
\begin{eqnarray}
\frac{\chi_{F,1}}{L} &=& \frac{10 \bar{M}-9}{48 \bar{M}^2}-\frac{\bar{M}^2-\bar{M}}{8 \bar{M}^2 \left(\sqrt{(\bar{M}-1) \bar{M}}-\bar{M} \tanh \left(\sqrt{(\bar{M}-1) \bar{M}}\right)\right)^2}\notag \\
&&+\frac{-12 \bar{M}^2+20 \bar{M}-21}{48 \bar{M} \sqrt{(\bar{M}-1) \bar{M}} \left(\sqrt{(\bar{M}-1) \bar{M}}-\bar{M} \tanh \left(\sqrt{(\bar{M}-1) \bar{M}}\right)\right)}-\frac{4 \bar{M}^3-8 \bar{M}^2+4 \bar{M}+3}{24 (\bar{M}-1)^2 \bar{M}^2}.
\end{eqnarray}

\textbf{Off-diagonal contribution to fidelity susceptibility.}
The off-diagonal matrix elements contribute to the fidelity susceptibility as follows
%%%%%%%%%%%%%%%%%%
\begin{equation}
\chi_{F,2} = -4 L \sigma(\bar M)  =  -4 L [\sigma_1(\bar M) + \sigma_{2}(\bar{M})]\label{eq:chiF2}
\end{equation}
where
%%%%%%%%%%%%%%
\begin{eqnarray}
\sigma_1(\bar M) &=& \frac{1}{4} \sum_{\bar k \in \mathcal E \backslash \lbrace 0 \rbrace} \sum_{\bar l \in \mathcal E \backslash \lbrace \pm \bar k \rbrace} \frac{\bar k^2 \bar l^2}{\mathcal D_{\bar k}\mathcal D_{\bar l} \left (\bar l^2 - \bar k^2\right )^2}  \\
\sigma_2(\bar M) &=& -\frac{1}{2} \sum_{\substack{\bar k \in \mathcal E^{+}\\ \bar l \in \mathcal E^{-}}} \frac{\bar k^2 \bar l^2}{\mathcal D_{\bar k}\mathcal D_{\bar l} \left (\bar l^2 - \bar k^2\right )^2}
\end{eqnarray}
%%%%%%%%%%%%%%%%%%%%
Here we introduced%\imEJK{Attention, I changed the notation of $\mathcal E$ as compared to notes.}
\begin{eqnarray}
\mathcal E^{+} &:=& \lbrace \bar k \in \mathbb C \vert f_{\rm even}(\bar k) = 0\rbrace,\\
\mathcal E^{-} &:=& \lbrace \bar k \in \mathbb C \vert f_{\rm odd}(\bar k) = 0\rbrace ,
\end{eqnarray}
with
\begin{equation}
f_{\rm even/odd} (\bar k) = \tan(\bar k/2) - \frac{\bar k}{\bar M \mp \sqrt{\bar k^2 + {\bar M}^2}}.
\end{equation}
The imaginary solution $\bar k = i\varpi$ solves $f_{\rm odd} (i \varpi L) = 0$ for $\bar M>1$ and is thus an element of $\mathcal E^{-}$.

It turns out that $\sigma_1$ can be solved in a closed form by taking a contour integral:
\begin{eqnarray}
\sigma_1(\bar{M}) &=& \frac{(\bar{M}-4) \bar{M} (2 \bar{M}-5)-15}{192 (\bar{M}-1)^2 \bar{M}^2}-\frac{\bar{M}-1}{32 \bar{M} \left(\bar{M} \tanh \left(\sqrt{\bar{M}^2-\bar{M}}\right)-\sqrt{\bar{M}^2-\bar{M}}\right)^2} \notag \\
&&+\frac{12 \bar{M}^2-20 \bar{M}+21}{192 \bar{M} \sqrt{(\bar{M}-1) \bar{M}} \left(\bar{M} \tanh \left(\sqrt{\bar{M}^2-\bar{M}}\right)-\sqrt{\bar{M}^2-\bar{M}}\right)}. \label{eq:sigma1}
\end{eqnarray}

we thus find that
\begin{equation}
\chi_{F,1} - 4 L \sigma_1 = 0, 
\end{equation}
in accordance with Eq.~\eqref{eq:partialcancellation}. The only contribution to the fidelity susceptibility stems from $\sigma_2(\bar M)$:
\begin{equation}
{ \frac{\chi_F}{L} = 2 \sum_{\substack{\bar k \in \mathcal E^{+}\\ \bar l \in \mathcal E^{-}}} \frac{\bar k^2 \bar l^2}{\mathcal D_{\bar k}\mathcal D_{\bar l} \left (\bar l^2 - \bar k^2\right )^2}.} \label{eq:SusceptiblitySum}
\end{equation}
We thus derived the result presented in Eq.~\eqref{eq:SuscSumMaintext} of the main text.

\begin{figure}
\includegraphics[scale=.5]{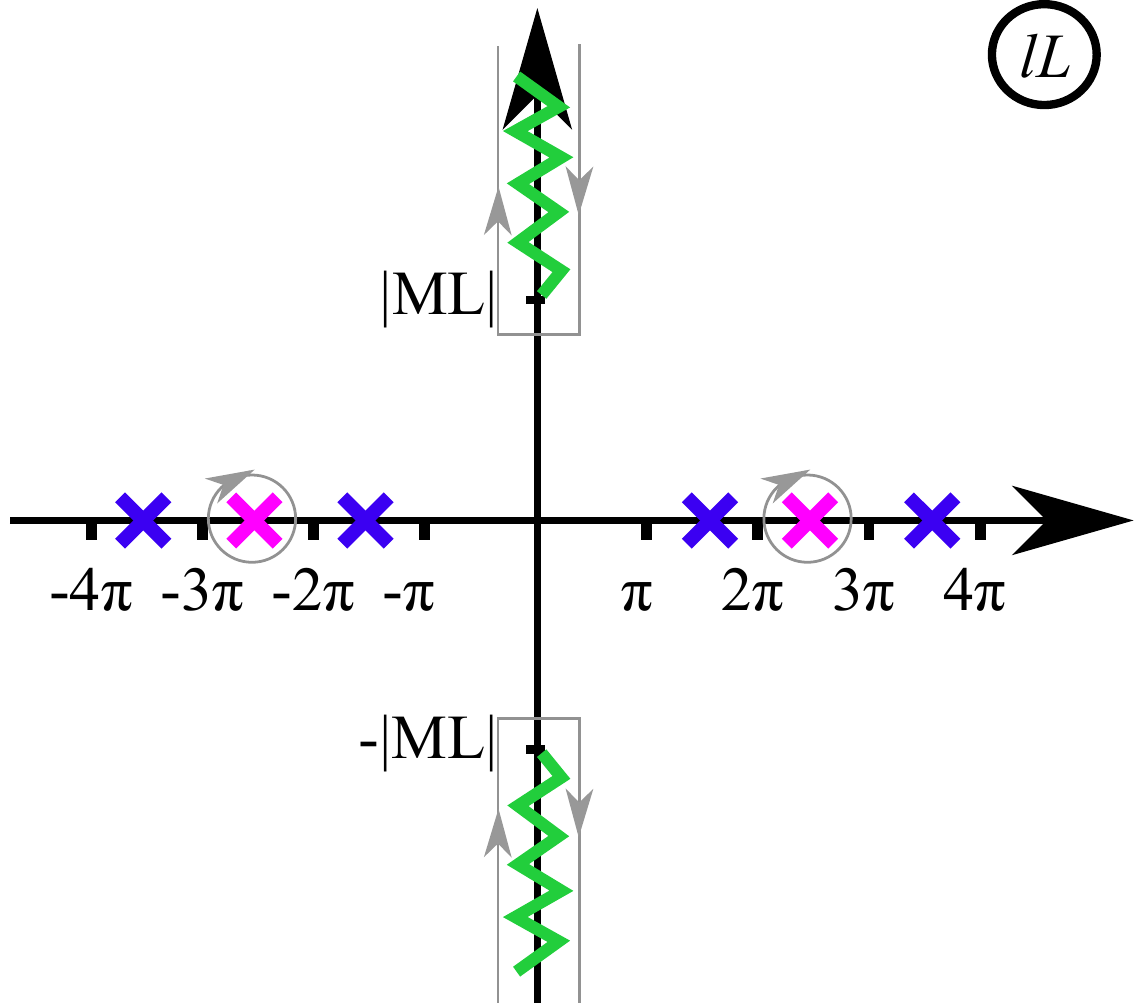} 
\hspace{2cm}
\includegraphics[scale=.5]{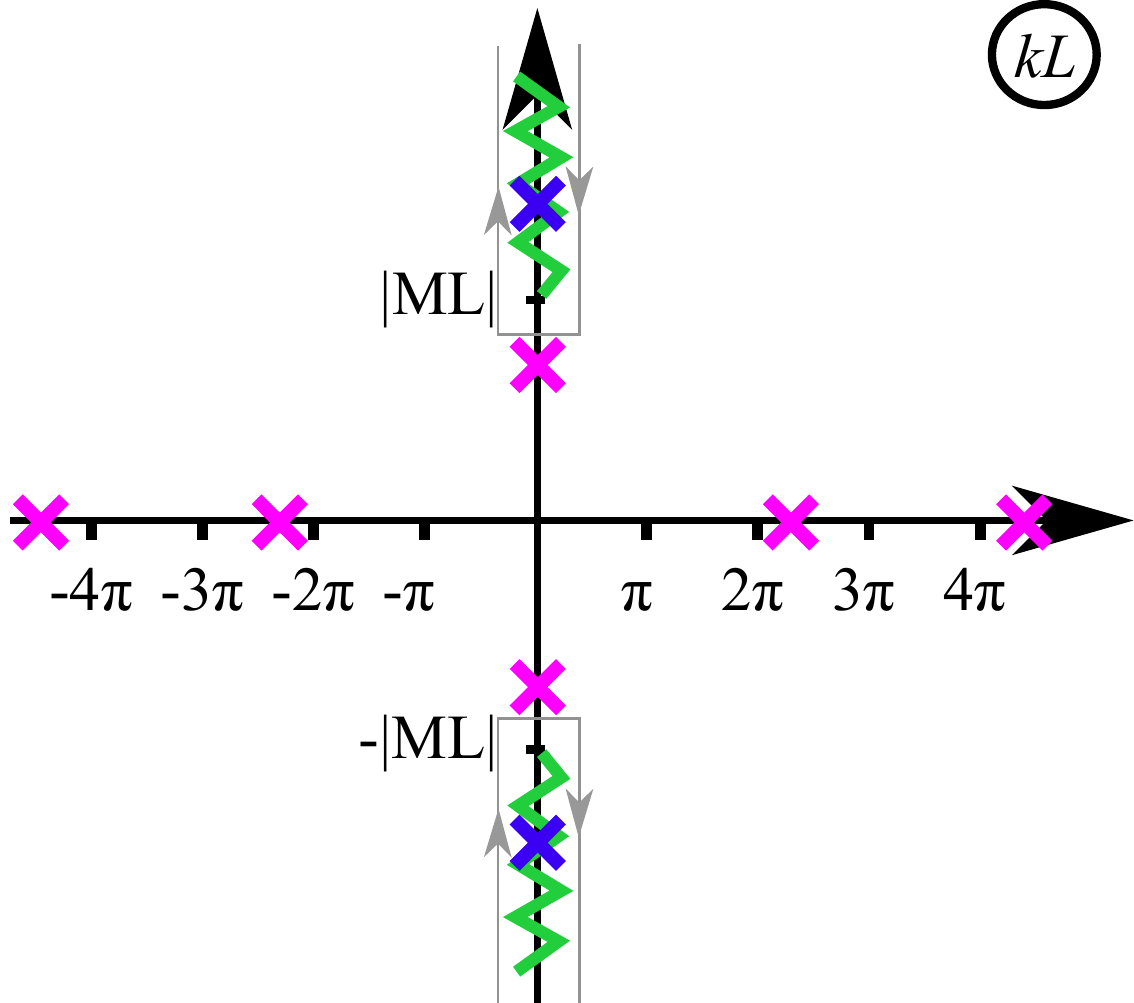} 
\caption{Integration contour for the evaluation of the sum $\sigma_2^{\rm even} (\bar k, \bar M)$ (left) and of the sum $\sum_{\bar k \in \mathcal E^{-}} \sigma_2^{\rm even}(\bar k, \bar M)$ (right). }
\label{fig:DoubleSumContour}
\end{figure}

\textbf{Evaluation of fidelity susceptibility.}
We proceed with the evaluation of the sum \eqref{eq:SusceptiblitySum} for the fidelity susceptibility. We use the following notation
\begin{equation}
\sigma_2(\bar M) = - \frac{1}{2}\sum_{\bar k \in \mathcal E^{+}} \sigma_2^{\rm odd}(\bar k, \bar M) - \frac{1}{2}\sum_{\bar k \in \mathcal E^{-}} \sigma_2^{\rm even}(\bar k, \bar M)
\end{equation}
For the contour integration, see Fig.~\ref{fig:DoubleSumContour}, we exploit the properties of the complex function 
\begin{eqnarray}
\tilde g_{\substack{\text{even}\\\text{odd}}} (\bar l) &=&   \frac{\mp \mathcal D_{\bar l}}{\sqrt{\bar l ^2 + {\bar M}^2} \tilde f_{\substack{\text{even}\\\text{odd}}}(\bar l) } = \frac{\mathcal D_{\bar l}}{2}\left ( \pm \frac{\tilde g_1(\bar l)}{\sqrt{\bar l^2 + {\bar M}^2}} + \tilde g_2(\bar l)\right ) \notag \\
&{\simeq}&\frac{ 1}{\bar l - \bar k_*}  \begin{cases}  \text{for } \bar l \rightarrow  \bar k_*, \;\bar k_*  \in  \mathcal E^{+} \text{ (upper sign),}\notag \\
\text{for } \bar l \rightarrow  \bar k_*, \;\bar k_*  \in  \mathcal E^{-}  \text{ (lower sign).}\end{cases}
\end{eqnarray}

Here we introduced
\begin{subequations}
\begin{eqnarray}
\tilde f_{\substack{\text{even}\\\text{odd}}}(\bar l) &=& (\bar M \mp \sqrt{\bar l^2 + {\bar M}^2})\tan(\bar l /2) - \bar l, \\
\tilde g_1 (\bar l) &=& \frac{1}{\bar l} + \frac{1}{\bar l \cos(\bar l) - {\bar M} \sin(\bar l)}, \\
\tilde g_2 (\bar l) &=& - \frac{\sin(\bar l)}{\bar l[ \bar l \cos(\bar l) - {\bar M} \sin(\bar l)]}.
\end{eqnarray}
\end{subequations}

This leads to
\begin{eqnarray}
\sigma_2^{\rm odd}(\bar k, \bar M) &=& \frac{1}{2} \sum_{\bar l \in \mathcal E^{-}} \frac{\bar k^2 {\bar{l}}^2}{\mathcal D_{\bar k}\mathcal D_{\bar l}(\bar k^2 - {\bar{l}}^2)^2} \notag \\
&=& \frac{\bar k^2 (\bar k^2 + {\bar M}^2 + 3 \bar M )}{16 \mathcal D_{\bar k} (\bar k^2 + {\bar M}^2)^2} +I_2(\bar k, \bar M)\\
\sigma_2^{\rm even}(\bar k, \bar M) &=&  \frac{1}{2}\sum_{\bar l \in \mathcal E^{+}} \frac{\bar k^2 {\bar{l}}^2}{\mathcal D_{\bar k}\mathcal D_{\bar l}(\bar k^2 - {\bar{l}}^2)^2} \notag \\
&=& \frac{\bar k^2 (\bar k^2 + {\bar M}^2 + 3 \bar M )}{16 \mathcal D_{\bar k} (\bar k^2 + {\bar M}^2)^2} -I_2(\bar k, \bar M).
\end{eqnarray}
Here, we introduced
\begin{equation}
I_2(\bar k, \bar M) = \frac{\bar k^2}{2 \mathcal D_{\bar k}}\frac{1}{\pi} \int_{\vert \bar M \vert}^\infty dx\, \frac{ x^2 }{ \sqrt{x^2-{\bar M}^2}} \frac{1}{x-\bar M \tanh (x)}  \frac{1}{\cosh (x)} \left (\frac{1}{\bar k^2+x^2} \right )^2.
\end{equation}

%%%%%%%%%%%%%%%%%%%%%%%%
%%%%%%%%%%%%%%%%%%%%%%%%

The second contour integration, cf. Fig.~\ref{fig:DoubleSumContour} (right), which is introduced to remove the $\bar k$ summation leads to the final expression for $\chi_F$ 
\begin{empheq}[box=\widefbox]{align}
\frac{\chi_F}{L}  &= \frac{-1}{16} \left \lbrace \frac{-1}{\bar M} + \frac{3+4\bar M}{{\bar M}^2(1-\tanh(\bar M))} - \frac{3}{(1-\tanh(\bar M))^2 {\bar M}^2} \right \rbrace \notag \\
& -\frac{2}{\pi^2} \lim_{\eta \rightarrow 0} \int_1^\infty \int_1^\infty \frac{du \, u^2}{\sqrt{u^2-1}}   \frac{dv \, v^2}{\sqrt{v^2-1}}  \frac{(u^2-v^2)^2-\eta^2}{[(u^2 - v^2)^2+\eta^2]^2} \notag \\
&\times \frac{1}{\cosh(\bar M u)[u-\tanh(\bar M u)]}\frac{1}{\cosh(\bar M v)[v-\tanh(\bar M v)]} . \label{eq:SSHIntegralSolution}
\end{empheq}
The second contribution, a principle value integral which stems from $I_2$, vanishes in the limit $ML \ll -1$. Due to the cancelling of exponential divergencies in the square bracket and the integral for $ML \rightarrow \infty$, it is numerically more stable to evaluate the fidelity by means of the sum, Eq.~\eqref{eq:SusceptiblitySum}. This is how the plots, Figs.~\ref{fig:SuscAllcases} and \ref{fig:ComparisonNumerics} were generated.

\subsection{Asymptotic behavior}

For the calculation of the asymptotic behavior of the quantum fidelity, we use the following approximate real solutions to Eq.~\eqref{eq:quantCond}
\begin{equation}
\bar p \stackrel{\vert \bar M\vert \gg 1}{\simeq} \pi n + \arctan{\frac{\pi n}{\bar M}}.
\end{equation}
For $M >0$ ($M <0$) $n$ is even for odd (even) states and odd for even (odd) states. We use that, for $\vert \bar M \vert \gg 1$ and any $n, m \in \mathbb N$, 
\begin{subequations}
\begin{eqnarray}
\arctan\left (\frac{ 2n \pi}{\bar M}\right ) + \arctan\left (\frac{ (2m-1)\pi}{\bar M}\right ) &\ll &  2n\pi+ (2m-1)\pi,  \\
\left \vert \arctan\left (\frac{ 2n\pi}{\bar M}\right ) - \arctan\left (\frac{ (2m-1)\pi}{\bar M}\right )\right \vert &\ll& \vert 2n\pi- (2m-1) \pi \vert,
\end{eqnarray}
\end{subequations}
so that we can expand the dispersive wave contribution to the sum Eq.~\eqref{eq:SusceptiblitySum} as follows
\begin{eqnarray}
\left .\frac{\chi_F}{L}  \right \vert_{\rm waves}&\simeq& 8 \sum_{n,m \in \mathbb N} \left  [\frac{\bar k^2 \bar l^2}{\mathcal D_{\bar k}\mathcal D_{\bar l} \left (\bar l^2 - \bar k^2\right )^2} + \frac{2 \bar k^2 \bar l \left(\bar k^2 \bar M^2+2 \bar l^4+\bar l^2 \bar M^2\right) \arctan\left(\frac{\bar l}{\bar M}\right)}{\left(\bar k^2-\bar l^2\right)^3 \left(\bar k^2+\bar M^2\right) \left(\bar l^2+\bar M^2\right)^2} + \bar l \leftrightarrow \bar k \right  ]_{\substack{\bar k = 2n \pi \\ \bar l = (2m -1) \pi}}
\end{eqnarray}
The first part of the sum is equivalent to Eq.~\eqref{eq:chi2SSHodd}. In the second term it is convenient to perform the $\bar k$ summation first, and then evaluate the sum over $\bar l$ as a Riemann integral. We keep only the leading and subleading contributions for $\bar M \rightarrow \infty$ and obtain
\begin{equation}
\left .\frac{\chi_F}{L}  \right \vert_{\rm waves} \simeq \frac{1}{16} \left (\frac{1}{\vert \bar M \vert} - \frac{4+\sgn{\bar M}}{4\bar M^2}\right ).
\end{equation}
For $\bar M \gg 1$, there is an additional contribution from the edge state
\begin{equation}
\left .\frac{\chi_F}{L}  \right \vert_{\rm edge + waves} \simeq 8 \bar M \sum_{m \in \mathbb N} \left  [\frac{\bar l^2}{\left (\bar l^2 +\bar M^2\right )^3} \right  ]_{\bar l = (2m-1) \pi } \simeq \frac{1}{4 \bar M^2}.
\end{equation}
This concludes the derivation of Eqs.~\eqref{eq:SSHevenAsymptotes} of the main text. Note that the behavior for $ML \ll -1$ can also be read off from Eq.~\eqref{eq:SSHIntegralSolution}.

%In contrast, 
%\begin{equation}
%\sigma_2(\bar{M}) = - 2\sum_{\bar k \in \mathcal E^{+}}\sigma_2'(\bar k, \bar{M})
%\end{equation}
%can not be solved in a closed form. We employ the following approximate solution
%
%%\begin{equation}
%%\sigma_2'(\bar k, {\bar M}) =  \underbrace{\frac{\bar k^2}{2 \mathcal D_{\bar k}}  \frac{1}{(\bar k^2 + {\bar M}^2)^2} \frac{\bar k^2 + {\bar M}^2 + 3 \bar M}{8} }_{=:\sigma_{2,\text{closed}}'(\bar k ,w ) }+  \underbrace{\frac{\bar k^2}{2 \mathcal D_{\bar k}}  \frac{1}{(\bar k^2 + {\bar M}^2)^2} \frac{1}{\cosh(\bar M)} I_2'(\bar k,{\bar M})}_{=:\sigma_{2,\text{Int.}}'(\bar k,w ) } 
%%\end{equation}
%%This result contains the following integral
%%\begin{subequations}
%%\begin{eqnarray}
%%I_2'(\bar k, {\bar M})  &=&\frac{1}{\pi} \int_{\vert \bar M \vert}^\infty dx\, \frac{ x^2 }{ \sqrt{x^2-{\bar M}^2}} \frac{1}{x(\bar M \tanh (x)}  \frac{\cosh (\bar M)}{\cosh (x)} \left (\frac{\bar k^2+{\bar M}^2}{\bar k^2+x^2} \right )^2 \notag \\
%%& \simeq& I_{2,a}'(\bar M) - \frac{2}{\bar k^2 + {\bar M}^2} I_{2,b}'(\bar M),% \label{eq:I2primeApprox}
%%\end{eqnarray}
%%where
%%\begin{eqnarray}
%% I_{2,a}'(\bar M) &=&\frac{1}{\pi} \int_{\vert \bar M \vert}^\infty dx\, \frac{ x^2 }{ \sqrt{x^2-{\bar M}^2}} \frac{1}{x(\bar M \tanh (x)}  \frac{\cosh (\bar M)}{\cosh (x)}  \\%\label{eq:I2aprimex} \\
%% I_{2,b}'(\bar M) &=&\frac{1}{\pi} \int_{\vert \bar M \vert}^\infty dx\, \frac{ (x^2 - {\bar M}^2) x^2 }{ \sqrt{x^2-{\bar M}^2}} \frac{1}{x(\bar M \tanh (x)}  \frac{\cosh (\bar M)}{\cosh (x)} \end{eqnarray}
%%\end{subequations}

\end{document}